\documentclass[twocolumn,preprintnumbers,amsmath,amssymb]{revtex4}
\usepackage{subfigure}
\usepackage{graphicx}
\usepackage{dcolumn}
\usepackage{bm}

\begin{document}

\title{Towards independent control of polar and azimuthal anchoring}
\author{C. Anquetil-Deck}
\email[Present address: Karlsruhe Institute of Technology, Institute for Meteorology and Climate Research, Atmospheric Aerosol Research Department (IMK-
AAF), Hermann-von-Helmholtz-Platz 1, D - 76344 Eggenstein-Leopoldshafen
Germany.]{}
\author{D.J. Cleaver}
\email{d.j.cleaver@shu.ac.uk}

\affiliation{ Materials and Engineering Research Institute, Sheffield Hallam
University, City Campus, Howard Street, Sheffield, S1 1WB,UK }

\author{J.P. Bramble}

\affiliation{ School of Physics and Astronomy, University of Leeds, Woodhouse Lane, Leeds, LS2 9JT, UK }

\author{T.J. Atherton}

\affiliation{ Department of Physics and Astronomy, Center for Nanoscopic Physics, Tufts University, Medford, MA 02155, USA}

\date{\today}

\begin{abstract}
Monte Carlo simulation, experiment and continuum theory are used to examine the anchoring exhibited by a nematic liquid crystal at a  patterned substrate comprising a periodic array of rectangles that, respectively, promote vertical and planar alignment. It is shown that the easy axis and effective anchoring energy promoted by such surfaces can be readily controlled by adjusting the design of the pattern. The calculations reveal rich behavior: for strong anchoring, as exhibited by the simulated system, for rectangle ratios $\geq 2$ the nematic aligns in the direction of the long edge of the rectangles, the azimuthal anchoring coefficient changing with pattern shape. In weak anchoring scenarios, however, including our experimental systems, preferential anchoring is degenerate between the two rectangle diagonals. Bistability between diagonally-aligned and edge-aligned arrangement is predicted for intermediate combinations of anchoring coefficient and system length-scale.
\end{abstract}

\maketitle

\section{Introduction}
Conventional uniform surface treatments for confined nematic liquid crystals (LCs),
such as rubbed or photoaligned polymers, are limited to a narrow specification:
they typically promote vertical or planar alignment where both the
preferred orientation and associated anchoring energy are difficult
to alter. In contrast, topographically or chemically patterned surfaces
permit essentially arbitrary control of the easy axis and anchoring
potential through appropriate adjustment of the pattern features~\cite{Harnau:2007p2047};
suitable patterning techniques include atomic force microscope scribing of polymer films~\cite{Kim_Nature_2002,Lee2004},
microcontact printing of self-assembled monolayers (SAMs)~\cite{Gupta:1996p1999,Gupta:1997p2000,RefWorks:13}
and photolithography~\cite{Bechtold2005,Bechtold:2005p3273,schadt92,stalder03}.
By imprinting a design of appropriate symmetry, it is also possible
to pattern a surface to promote more than one stable alignment orientation~\cite{Baek-woonLee03302001,Yi:2008p2678,Kim2001,Kim_Nature_2002,yoneya2002}
thus enabling the fabrication of bistable devices~\cite{ZBD,bryanbrown97,Kim_Nature_2002,Kim2003,kitson02,stalder03}.
Very recently, SAM-patterned substrates have been used to achieve rapid switching~\cite{NEWPAPER}.
\\[0.5cm]
The purpose of the present paper is to investigate the alignment
behavior of a nematic confined by patterned surfaces decorated with a stretched-chessboard-like
array of rectangles that alternately promote planar and vertical alignment.
This arrangement is of interest since it is intermediate between arrays of stripes and squares, patterns
which have previously been shown to promote qualitatively different anchoring behaviors.
\\[0.5cm]
For striped systems, it was found
experimentally by Lee and Clark~\cite{Baek-woonLee03302001} that the
polar orientation of the bulk nematic depends on the relative widths
of the vertical- and planar-promoting stripes while the azimuthal alignment consistently lies
along the lengths of the stripes. It was proposed that the latter was due
to elastic anisotropy, i.e. the differing energetic costs of various
symmetries of bulk orientational deformation. Calculations using continuum theory~\cite{barbero92,Atherton:2006p31}
and Monte Carlo (MC) simulation of hard particles~\cite{Bramble2007}
affirmed this notion and showed that it held
across the very different length scales over which these two theoretical techniques apply.
Further, it was shown that by adjusting the relative stripe width and cell thickness,
the polar anchoring angle could be altered continuously from planar
to vertical~\cite{Atherton:2006p31,Deck2010}. Despite
the apparent simplicity of the striped pattern, in some scenarios its
phase behavior is further enriched: if two striped substrates sandwich a cell that is only a few molecular
lengths thick, the nematic may form separate vertical and planar domains
``bridging'' the film~\cite{Deck2010}. Alternatively,
if some of the stripes are sufficiently narrow, the pattern is neglected
by the structure within the nematic~\cite{Atherton:2010p3029,Kondrat2003}.
\\[0.5cm]
As expected from symmetry~\cite{Yi:2008p2678,Kim_Nature_2002},
degenerate alignment is observed for systems comprising chessboard-like arrangements of squares promoting competing alignments. Experimental studies on SAM-based square-patterned systems have shown that, for feature sizes of
$\simeq 30$ microns, the favored alignment runs along the diagonals of the planar-promoting squares~\cite{Bramble2007}.
When the patterning length-scale is reduced to that accessible to particle-based simulation, however, the preferred orientations
run along the pairs of opposite edges of the squares~\cite{Deck2012}. An anchoring transition between these arrangements is
then predicted for, e.g., a sufficiently weak polar anchoring condition. Unlike the striped systems, however, no
``bridging'' behavior has been observed for square-patterned systems, even for very thin films.
\\[0.5cm]
Here, we investigate the way in which the strong azimuthal coupling
and variable polar anchoring associated with nematics on striped substrates segues into
the anchoring transition found for square-patterned substrates. We do this by
studying the polar and azimuthal anchoring behaviors accessible to
substrate patternings based on rectangles of adjustable length-scale and
aspect ratio. This is achieved through a combination of molecular-level MC simulation,
experiment and continuum theory.
The paper is organized as follows: the MC simulation methodology is described
in section II and associated results are presented in section III.
Some preliminary experimental results are then given in section
IV and a continuum calculation is performed and used to reconcile the preceding results,
in section V. Conclusions are drawn in section VI.

\section{Monte Carlo Model and Simulation Details}
\begin{figure}[!h]
\begin{center}
\includegraphics[scale=0.250]{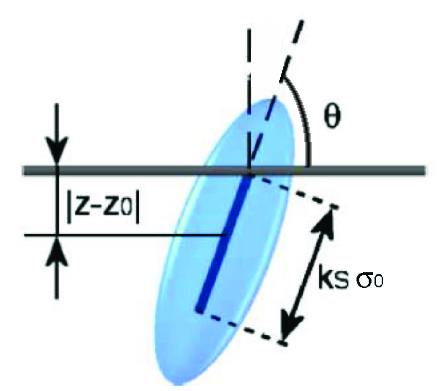}
\caption{(Color online)Schematic representation of the geometry used for the
hard needle-wall (HNW) particle-substrate interaction
\cite{Cleaver_Teixeira}.} \label{hnw}
\end{center}
\end{figure}

\begin{figure}[!h]
\begin{center}
\includegraphics{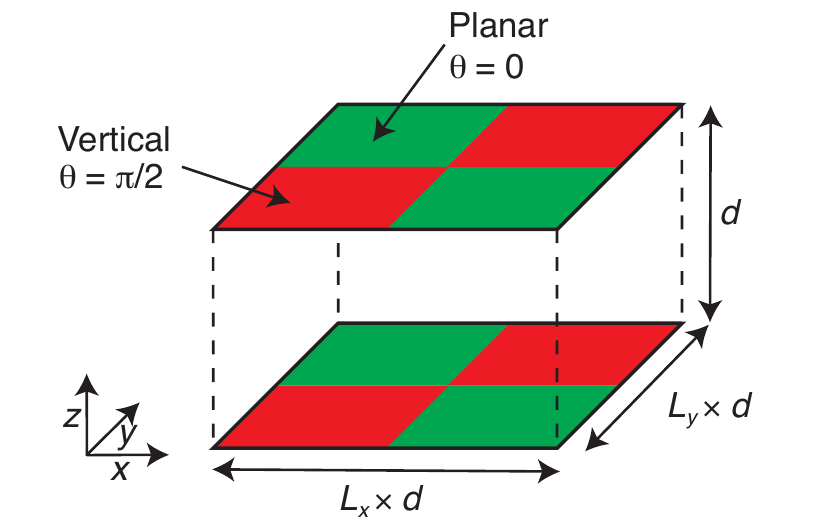}
\caption{(Color online)Schematic representation of rectangle
patterned systems tiled with vertical-inducing (red/dark) and planar-inducing (green/light) substrate regions. The azimuthal angle $\phi$ is zero along the $x$-axis} \label{schema_systems}
\end{center}
\end{figure}

To initiate this study, we have performed a series of MC simulations of
rod-shaped particles confined in slab geometry between two planar
walls. The model used is essentially that described in~\cite{Deck2012}. Briefly, inter-particle interactions have been modelled through the
Hard Gaussian Overlap (HGO) potential~\cite{Kike} which can be seen as equivalent to the well known
Gay-Berne model~\cite{Gay_berne}, stripped of its attractive interaction. The particle-substrate interactions have been modelled using the hard
needle-wall potential (HNW)~\cite{Cleaver_Teixeira} where a hard axial needle
of length $\sigma_{0}k_{s}$ is placed at the centre of each particle
(see Figure ~\ref{hnw}).
The parameter $k_{s}$ provides a molecular-level control on the surface anchoring properties.
The results presented in Sections \ref{Rect1} and \ref{Rect2}
have been obtained for systems of 864 HGO particles of length to breadth ratio $\kappa = 3$,
confined between two rectangle patterned substrates.
The substrates were separated by a distance $L_{z}$ =
4$\kappa\sigma_{0}$, where $\sigma_0$ is the particle diameter, periodic boundary conditions being imposed
in the $x$- and $y$-directions.
\\[0.5cm]
On each substrate, $k_{s}$ was set to a
vertical-aligning value ($k_{s}\leq 1.0$) for a portion of its area and a planar
value ($k_{s}\geq 2.0$) for the remainder and
sharp boundaries were imposed between the competing alignment
regions. The patterns on the top and bottom surfaces
have been kept in perfect registry with one another, as shown
in the schematic Fig.~\ref{schema_systems}.
\\[0.5cm]
Each system has
been initialized at low density and gently compressed by
decreasing the box dimensions $L_{x}$ and $L_{y}$ whilst keeping the substrate separation $L_{z}$
and the rectangular ratio $R=L_{x} / L_{y}$ fixed. At each density,
run lengths of 1 million MC sweeps (where one sweep represents one
attempted move per particle) were performed, averages and profiles
being accumulated for the final 500 000 sweeps.
\\[0.5cm]
Analysis has been performed by dividing stored system
configurations into 100 equidistant constant-$z$ slices and
performing averages of relevant observables in each slice. This
yields profiles of quantities such as number density, $\rho^{*}(z)$, from which
structural changes can be assessed. Orientational order profiles
have also been calculated, particularly
\begin{eqnarray}
Q_{zz}(z)= \frac{1}{N(z)} \sum_{i=1}^{N(z)}\bigg (\frac{3}{2}
u_{i,z}^{2}-\frac{1}{2}\bigg )
\end{eqnarray}
which measures variation across the confined films of orientational
order measured with respect to the substrate normal. Here $N(z)$ is
the instantaneous occupancy of the relevant slice. We have also further
subdivided the system to assess
lateral inhomogeneities induced by the patterning. Specifically, we have computed profiles corresponding to particles residing between the vertical-aligning and planar-aligning substrate regions.

\section{Simulation Results}
\subsection{Influence of the surface interaction parameter}
\label{Rect1}
In this section, we assess the influence of microscopic rectangular patterning on the structure and anchoring of a confined LC film. To this end, we present results obtained from full compression sequences of MC simulations performed on a series of systems with differing pairs of surface interaction parameters. For reasons of space we only show data for high density ($\rho=0.4$) systems, and concentrate on the influence of the substrate parameters.

\begin{figure}[!h]
\begin{center}
\subfigure[\label{fig:rect2p}
]{\includegraphics[scale=0.16] {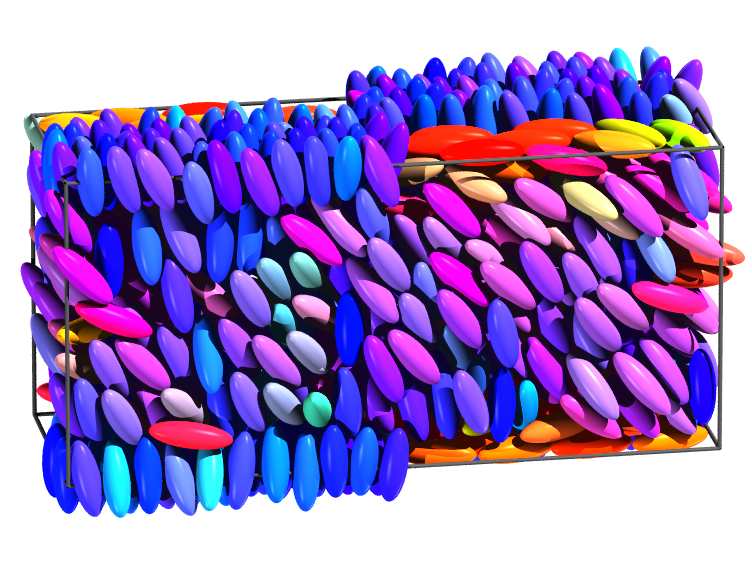}}
\subfigure[\label{fig:rect6p}
]{\includegraphics[scale=0.16] {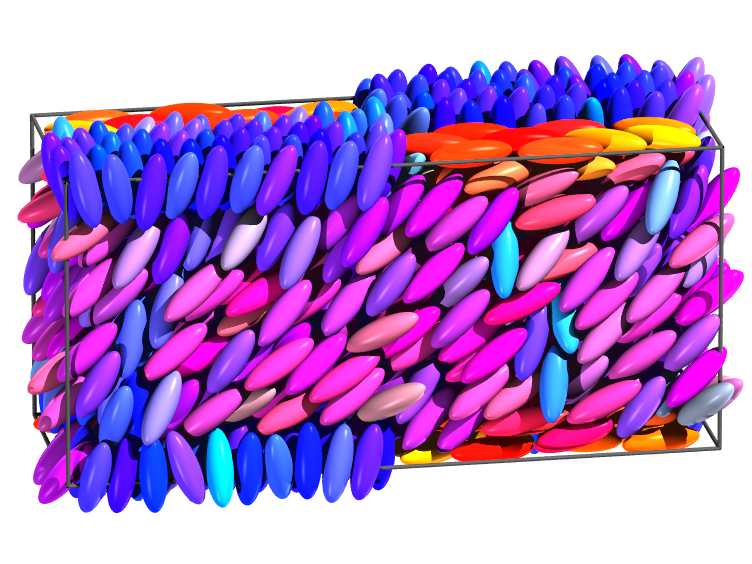}}
\subfigure[\label{fig:rect4p}
]{\includegraphics[scale=0.16] {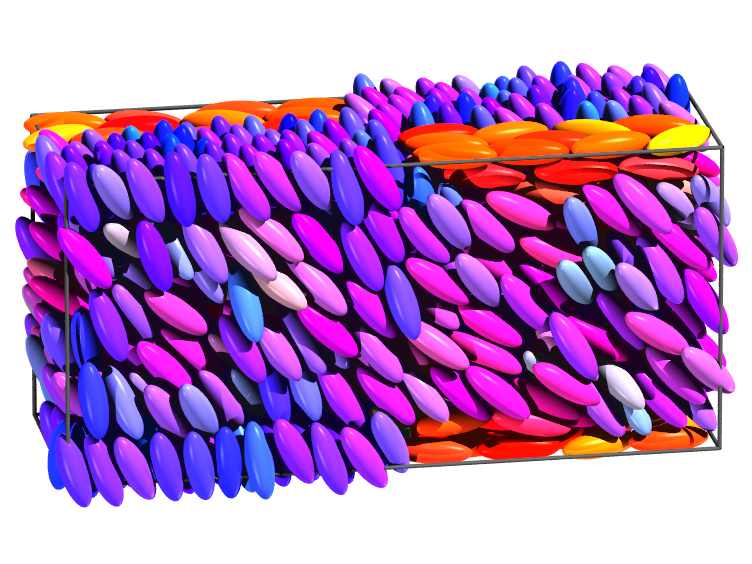}}
\subfigure[\label{fig:rect10pp}
]{\includegraphics[scale=0.16] {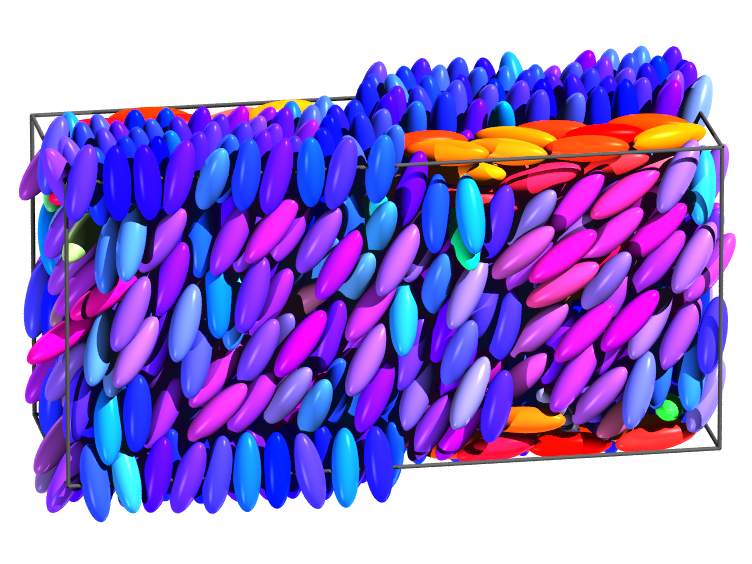}}
\subfigure[\label{fig:rect8pp}
]{\includegraphics[scale=0.16] {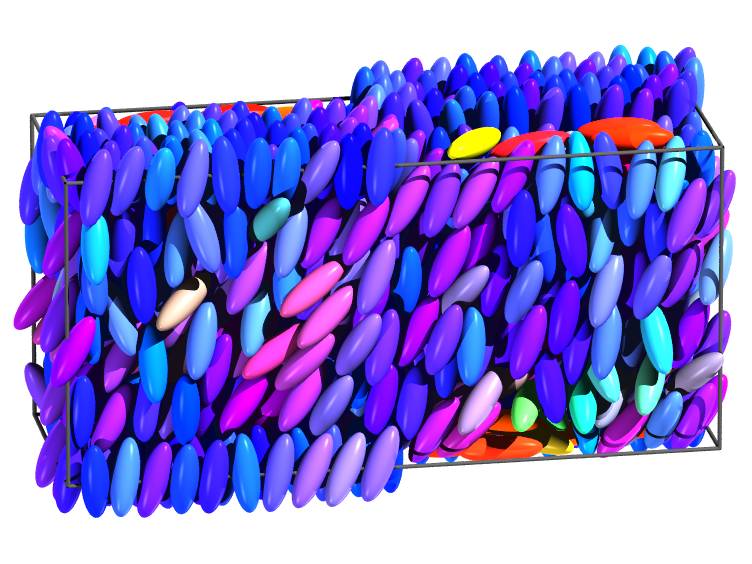}}
\caption{\label{snapshot_rectangle_influenceratio_Lz=4K1}(Color online)Snapshots
of system with $L_{x}$=3$L_{y}$ and different combinations of planar and vertical $k_{s}$ values. (a) $k_{s}$=0.5; $k_{s}$=3; (b) $k_{s}$=0; $k_{s}$=3; (c) $k_{s}$=1; $k_{s}$=3; (d) $k_{s}$=0; $k_{s}$=2.5; (e) $k_{s}$=0; $k_{s}$=2 }
\end{center}
\end{figure}

We initially consider systems for which the rectangular ratio $R=3$. We first used a combination of strong planar alignment of the
molecules on the substrate ($k_{s}=3$) with a strong vertical
alignment ($k_{s}=0$). Then, we slightly weakened the vertical
alignment ($k_{s}=0.5$) before going on to use a weak vertical
alignment ($k_{s}=1$). The high density snapshots corresponding to
such systems are represented in Fig.
~\ref{snapshot_rectangle_influenceratio_Lz=4K1}. These clearly indicate that,
for all combinations of $k_{s}$ values, these systems exhibited a central ordered  monodomain at high density.
From these snapshots it is apparent that the monodomains were all aligned in the $x$-$z$-plane. This is consistent with what was observed for equivalent stripe-patterned systems~\cite{Deck2010}. It suggests, then, that
for an edge ratio of 3, rectangle patterned substrates
act rather like well defined stripes, i.e., that it is straightforward to pin the azimuthal
angle using the shape asymmetry of rectangular substrate patterns. The
second item of note from these snapshots is that the surface
patterns do not `bridge' across the film. As with square patterns~\cite{Deck2012},
this is presumably due to the different interfacial stabilities that would pertain at
the resultant twist-like and bend-like domain boundaries.
\\[0.5cm]
\begin{figure}[!h]
\begin{center}
\subfigure[\label{fig:rectcomp2_6_4_homeo}
]{\includegraphics[width=0.5\textwidth]
{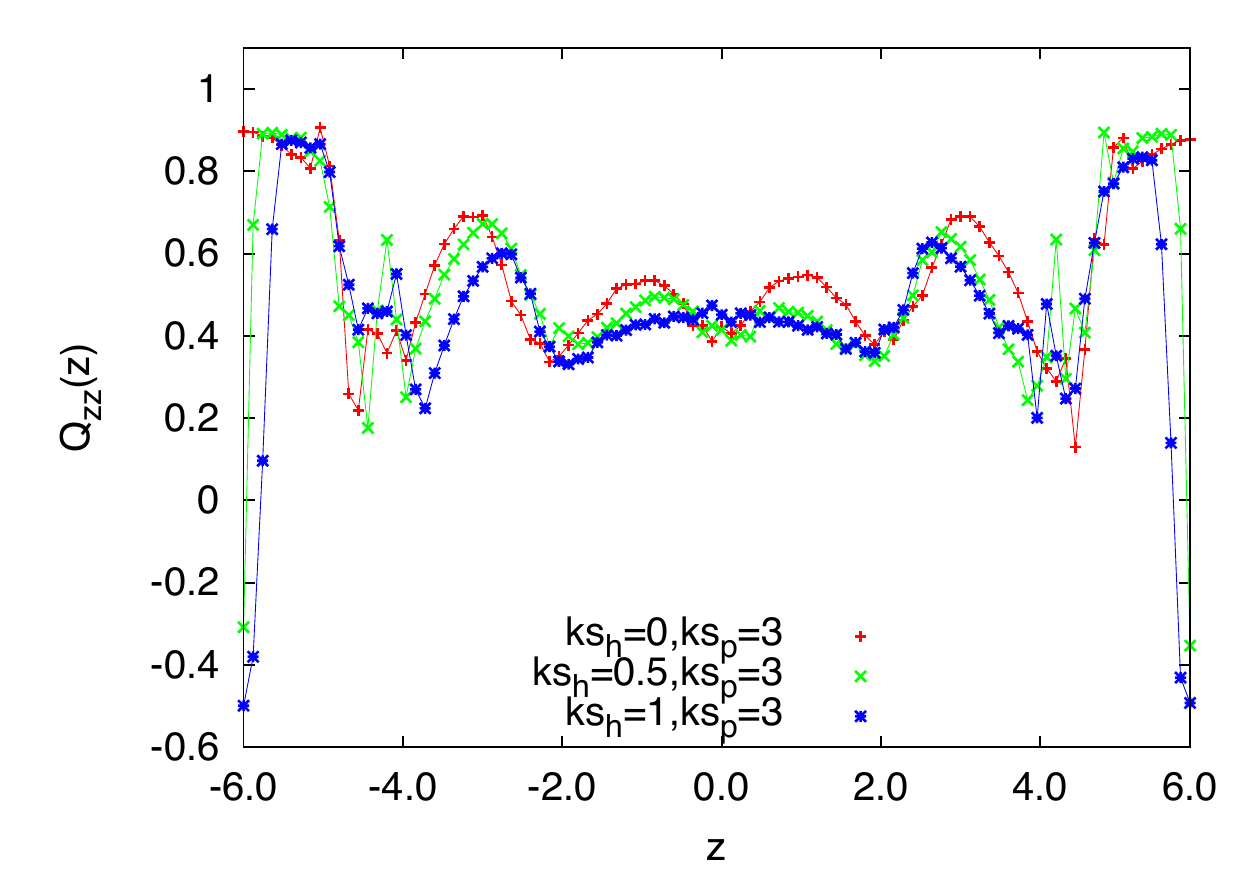}}
\subfigure[\label{fig:rectcomp2_6_4_planar}
]{\includegraphics[width=0.5\textwidth]
{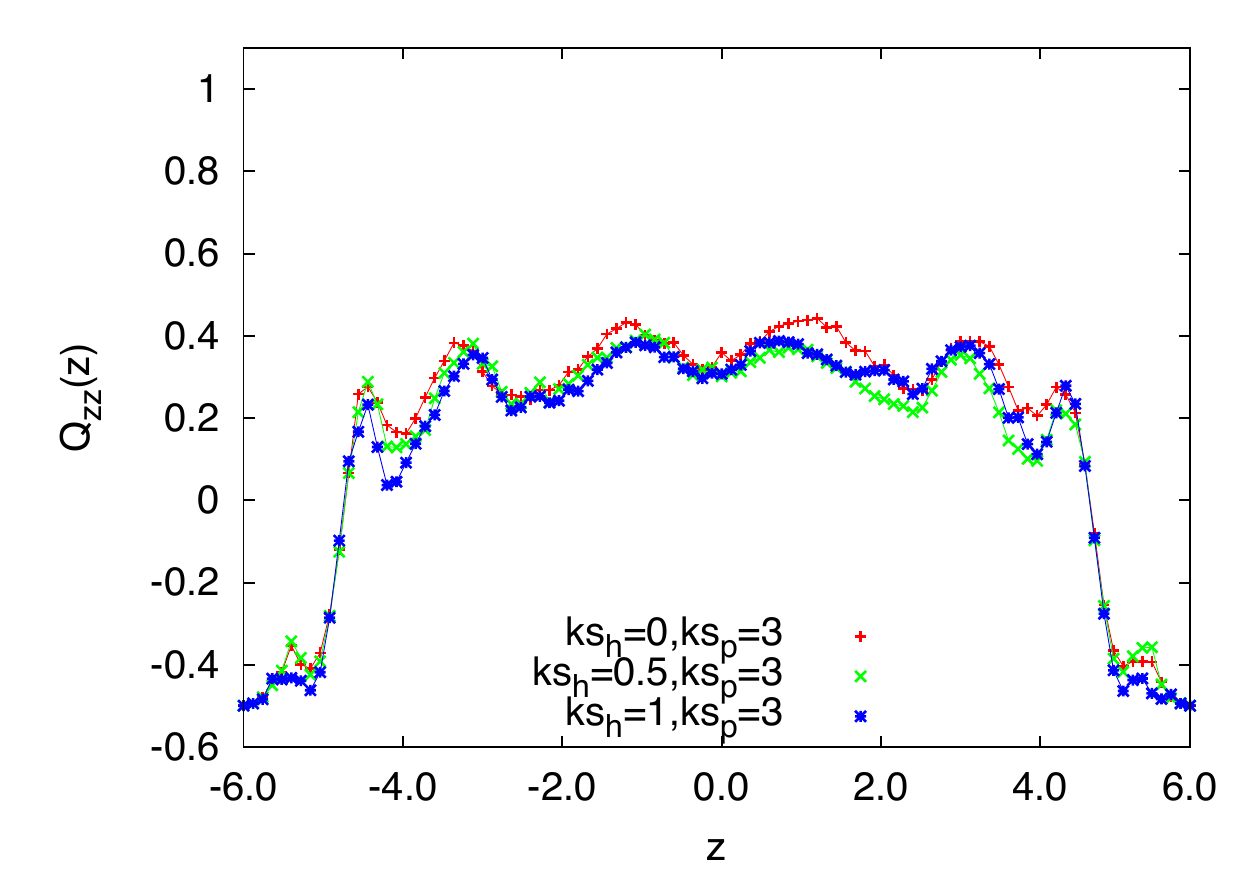}}
\caption{\label{fig:rectcomp2_6_4}(Color online)Influence of $k_{s}$
on the $\rho=0.4$ $Q_{zz}$ profiles for the differently-anchored subsystems with rectangle ratio $R$=3 and strong planar anchoring
(a) between vertically-aligned substrate regions; (b) between planar-aligned substrate regions}
\end{center}
\end{figure}
\begin{table}[h]
\centering

\begin{tabular}{ccccc}
\hline
 &\multicolumn{2}{c}{Planar} &\multicolumn{2}{c}{Vertical}\\
$L_{x}/L_{y}$ & $k_{s}$ & $\theta_{z}$($^{\circ}$) & $k_{s}$ & $\theta_{z}$($^{\circ}$)\tabularnewline
\hline
\hline
3 & 3 & 53 & 0 & 59\\
3 & 3 & 50 & 0.5 & 57\\
3 & 3 & 50 & 1 & 56\\
3 & 2.5 & 60 & 0 & 64\\
3 & 2 & 65 & 0 & 68\\
2 & 3 & 55 & 0 & 61\\
2 & 3 & 54 & 0.5 & 59\\
2 & 3 & 45 & 1 & 50\\
2 & 2.5 & 61 & 0 & 66\\
2 & 2 & 75 & 0 & 75\\
\hline
\end{tabular}

\caption{Average tilt angle, by surface region, for different rectangle ratios $L_{x}$/$L_{y}$ and substrate couplings $k_{s}$.}
\label{tiltrect3}
\end{table}

\begin{figure}[!h]
\begin{center}
\subfigure[\label{fig:rectcomp2_10_8_homeo}
]{\includegraphics[width=0.5\textwidth]
{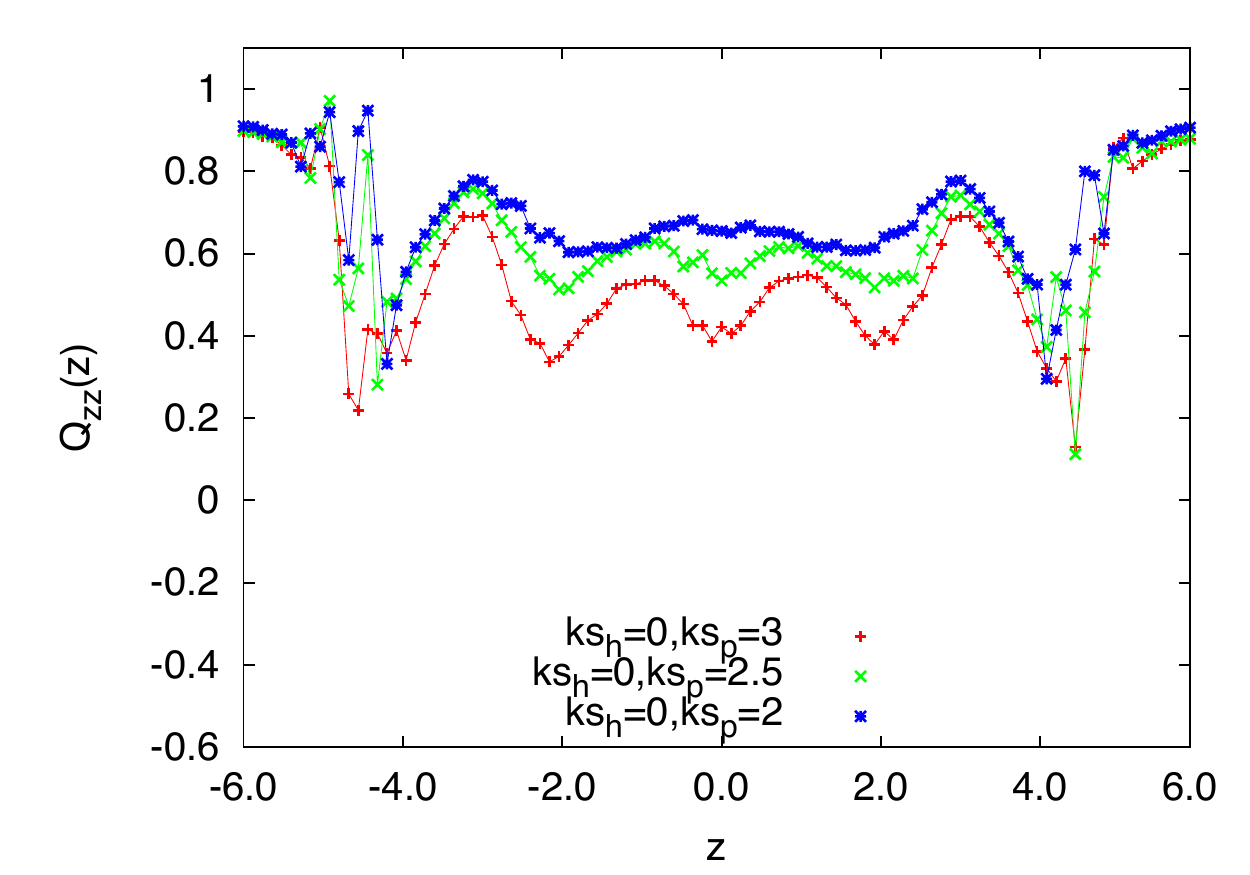}}
\subfigure[\label{fig:rectcomp2_10_8_planar}
]{\includegraphics[width=0.5\textwidth]
{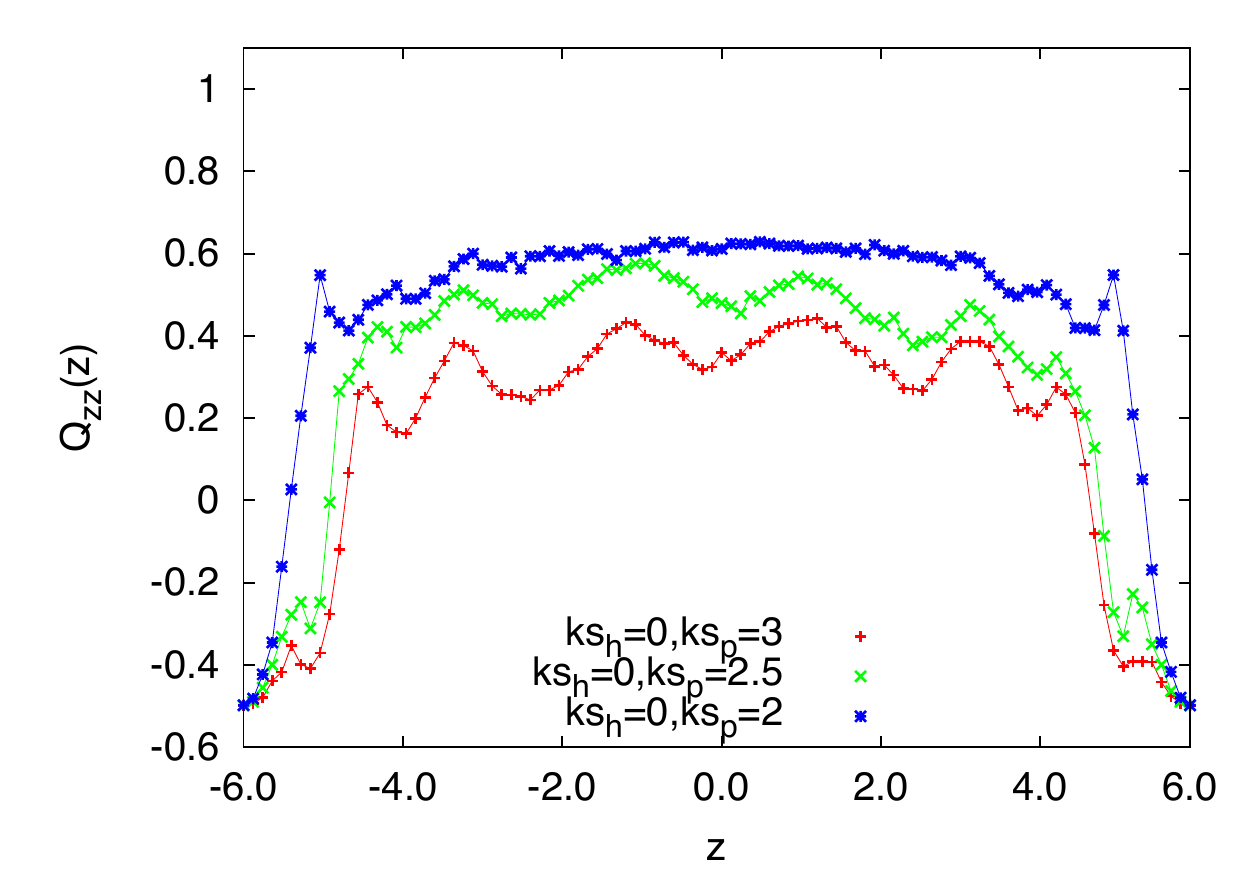}}
\caption{\label{fig:rectcomp2_10_8}(Color online)Influence of $k_{s}$
on the $\rho=0.4$ $Q_{zz}$ profiles for the differently-anchored subsystems with rectangle ratio $R$=3 and strong vertical anchoring
(a) between vertically-aligned substrate regions; (b) between planar-aligned substrate regions.}
\end{center}
\end{figure}

To substantiate this assessment, we plot, in Fig. ~\ref{fig:rectcomp2_6_4}, the
corresponding $Q_{zz}$ profiles to quantify the differences
between these three systems with strong planar anchoring (i.e. $k_{s}=3$).
From these it is apparent that in the vertical regions,
the $Q_{zz}$ value is slightly greater for strong vertical parameterizations (see Fig.
~\ref{fig:rectcomp2_6_4_homeo}). This difference is not apparent,
though, in the planar regions (see Fig.
~\ref{fig:rectcomp2_6_4_planar}).
Because of the small differences in
these graphs, we have also calculated the
average bulk tilt angles. These are presented in Table
~\ref{tiltrect3} and confirm that these systems exhibit very
similar tilt angles in the differently anchored regions,
typical variations being only $7^\circ$ within each system. These modest variations correspond to
nematic monodomains with small undulations of the polar anchoring orientation. As expected, on decreasing
the strength of the vertical anchoring at the surface,
the bulk alignment becomes increasingly planar: the tabulated values for $\theta_{z}$ decrease
with increase in the vertical-region $k_{s}$ value. Whilst this variation is very weak,
the presence of the vertical-aligning surface regions is substantial; the measured tilt angles are far from the
$0^{\circ}$ that would be seen in the absence of these vertical-aligning substrate regions.
\\[0.5cm]
We next consider the influence of the strength of the planar anchoring regions in the presence of strong vertical anchoring
regions. This is achieved by simulating $k_{s}=0$ regions combined with strong planar ($k_{s}=3$), moderate planar ($k_{s}=2.5$) and weak planar($k_{s}=2$) regions, respectively. From the variation of the corresponding $Q_{zz}$ profiles (see Figs.
~\ref{fig:rectcomp2_10_8}), it is apparent that decreasing the
planar anchoring strength, causes the $Q_{zz}$ profiles to become more positive throughout
the confined film. As confirmed by the snapshots (Fig.~\ref{snapshot_rectangle_influenceratio_Lz=4K1})
and the corresponding tilt angle data in
Table~\ref{tiltrect3}, this means that the central domains become increasingly aligned
perpendicular to the surfaces. Again, this makes intuitive sense, since
the relative influence of the vertical anchoring contribution would be
expected to grow in such a circumstances.
The bulk tilt angle is, though, more sensitive here to change of the planar-region coupling parameter than
it was previously to change of the vertical-region coupling parameter.

\subsection{Influence of the rectangle ratio $R=L_{x}/L_{y}$}
\label{Rect2}
\begin{figure}[!h]
\begin{center}
\subfigure[\label{fig:rect1pp}
]{\includegraphics[scale=0.16] {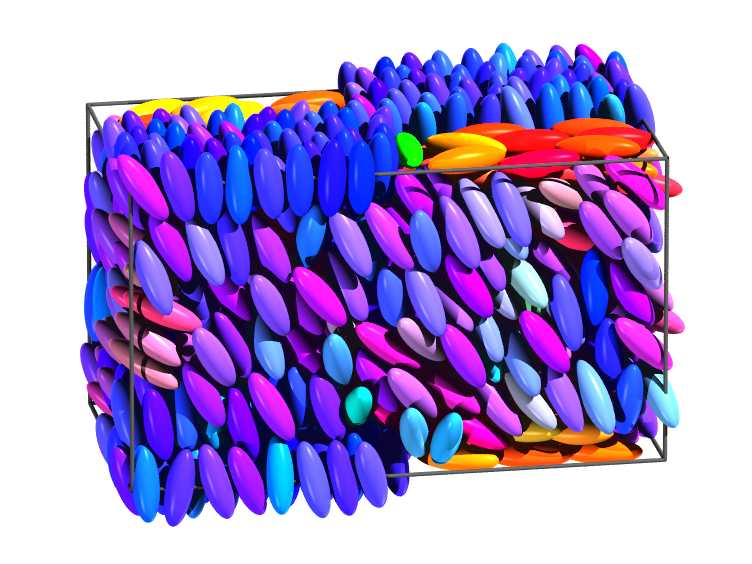}}
\subfigure[\label{fig:rect5p}
]{\includegraphics[scale=0.16] {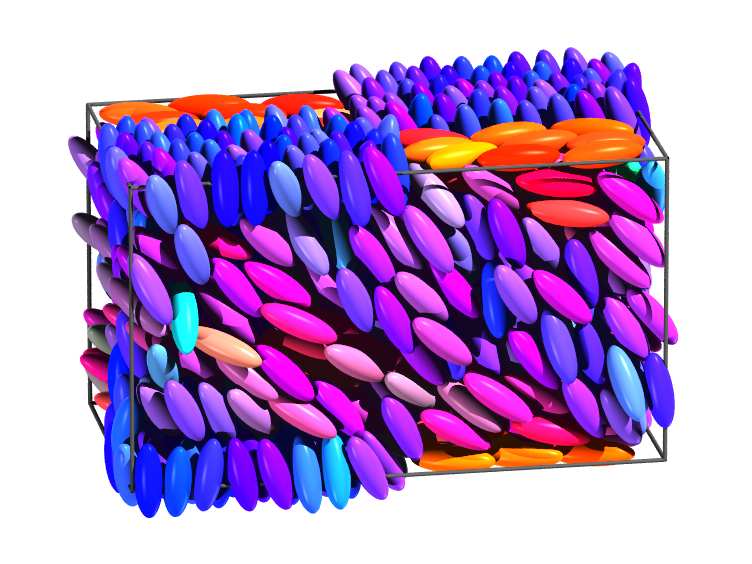}}
\subfigure[\label{fig:rect3p}
]{\includegraphics[scale=0.16] {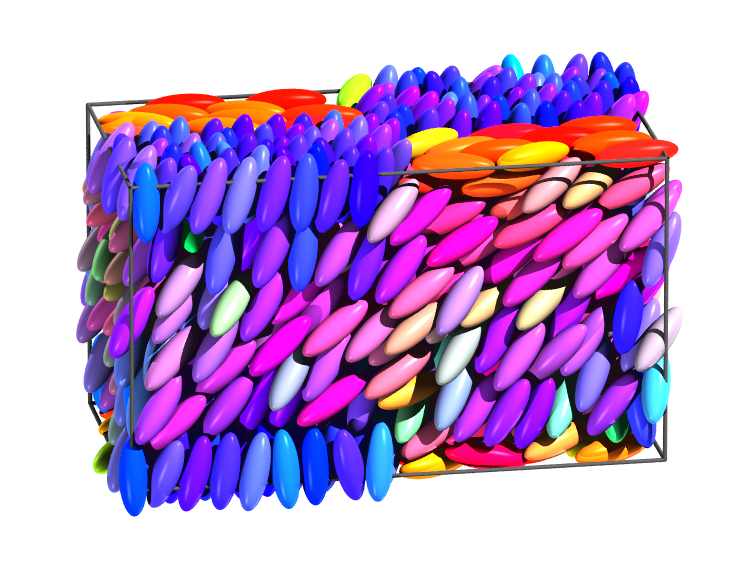}}
\subfigure[\label{fig:rect9pp}
]{\includegraphics[scale=0.16] {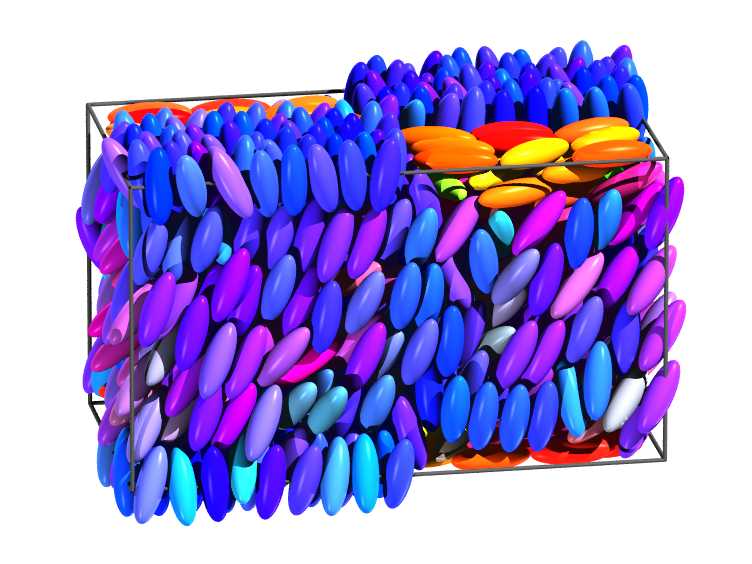}}
\subfigure[\label{fig:rect7pp}
]{\includegraphics[scale=0.16] {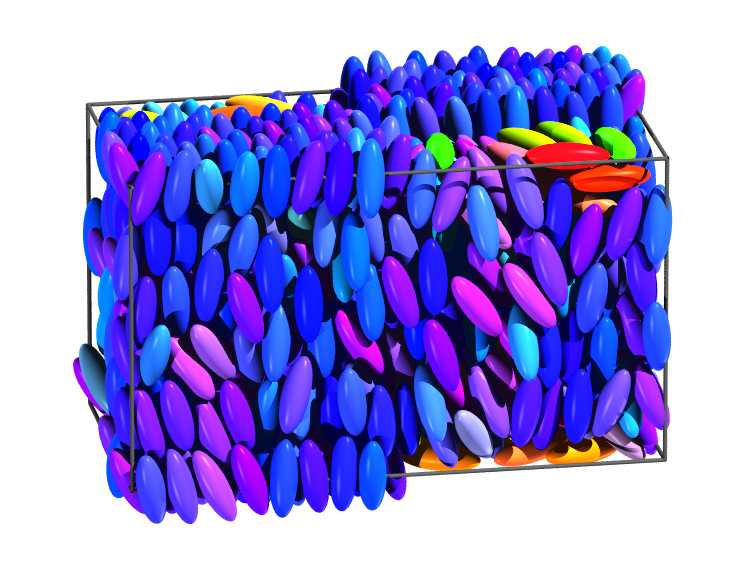}}
\caption{\label{snapshot_rectangle_influenceratio_Lz=4K2}(Color online)Snapshots
of system with $L_{x}$=2$L_{y}$ and different combinations of planar and vertical $k_{s}$ values.
Strong vertical anchoring. (a) $k_{s}$=0; $k_{s}$=3; (b) $k_{s}$=0.5; $k_{s}$=3; (c) $k_{s}$=1; $k_{s}$=3; (d) $k_{s}$=0; $k_{s}$=2.5; (e) $k_{s}$=0; $k_{s}$=2.}
\end{center}
\end{figure}

\begin{figure}[!h]
\begin{center}
\subfigure[\label{fig:rectcomp1_9_7_homeo}
]{\includegraphics[width=0.5\textwidth]
{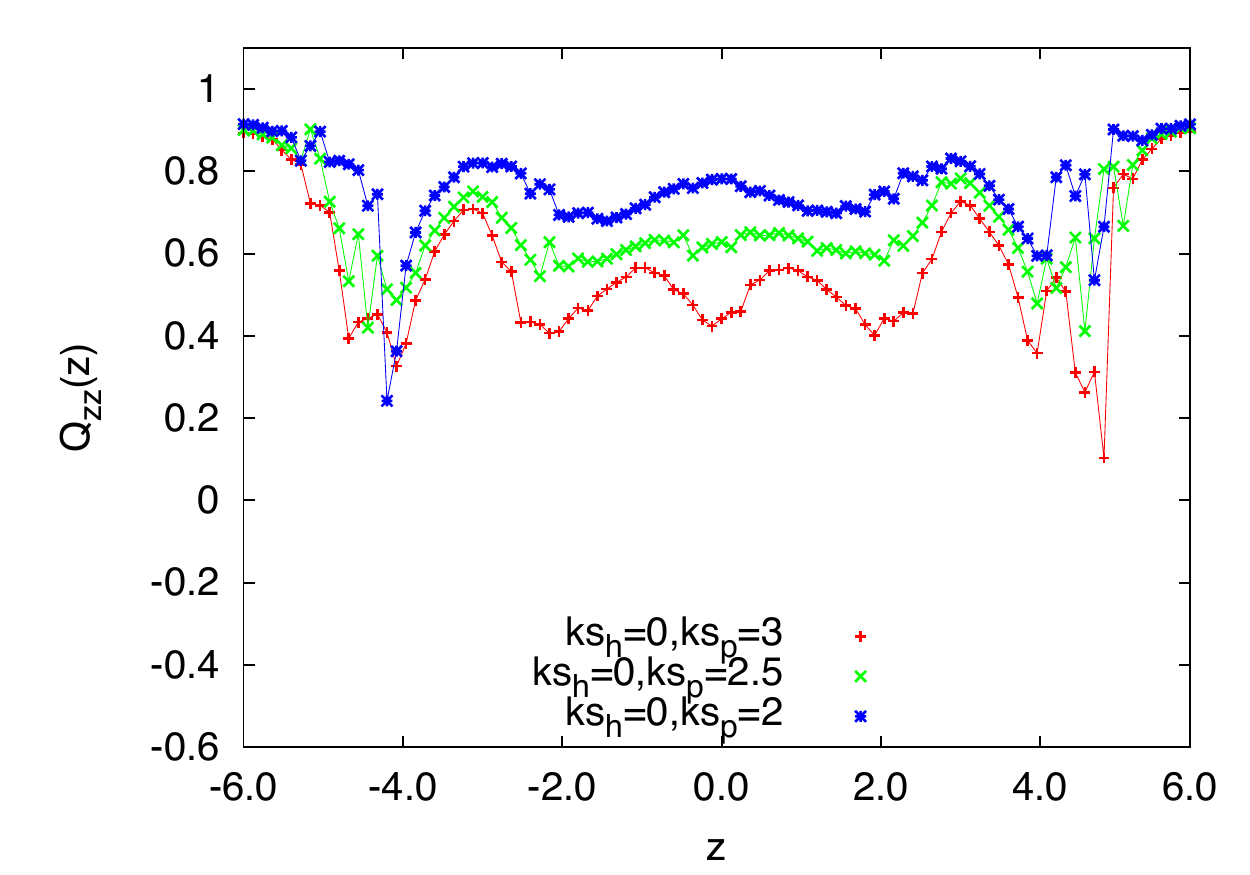}}
\subfigure[\label{fig:rectcomp1_9_7_planar}
]{\includegraphics[width=0.5\textwidth]
{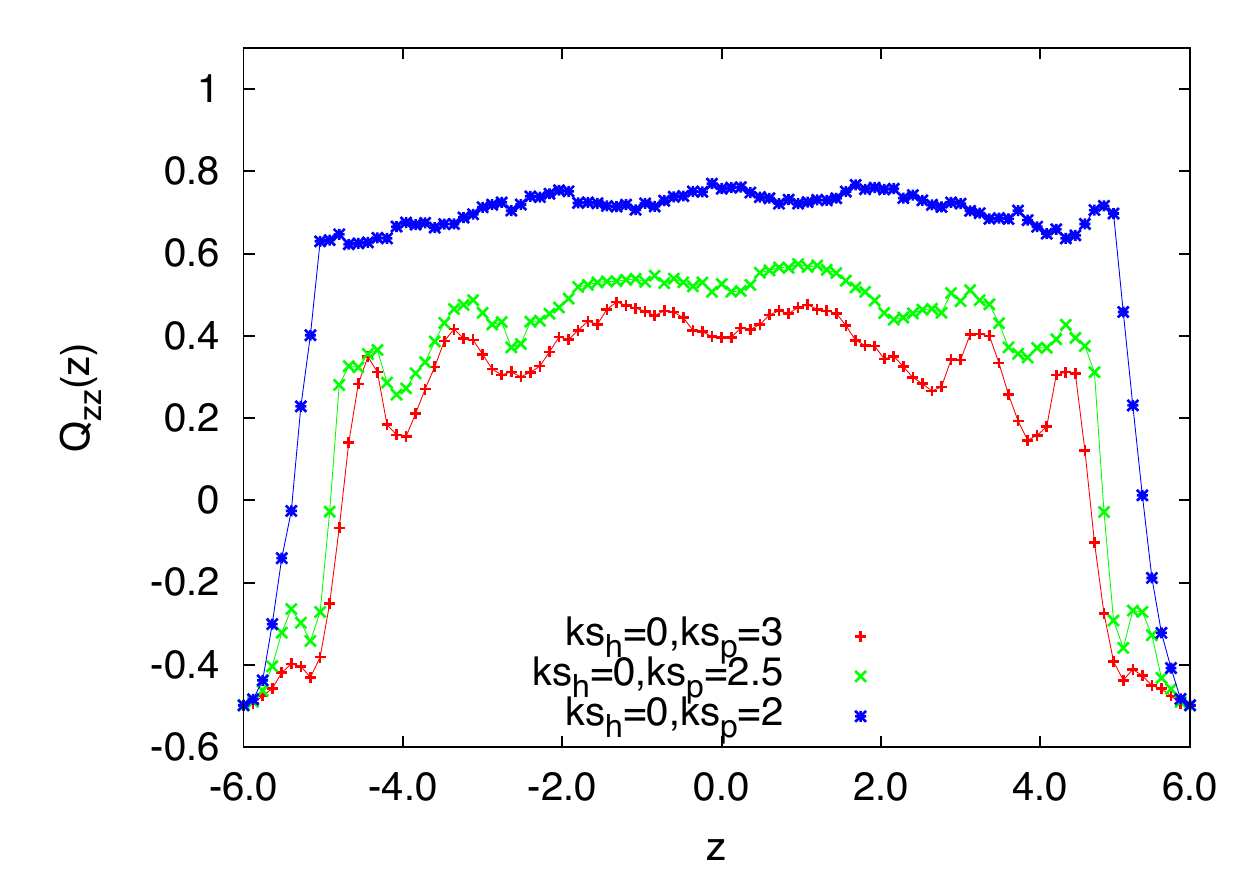}}
\caption{\label{fig:rectcomp1_9_7} (Color online)Influence of $k_{s}$
on the $\rho=0.4$ $Q_{zz}$ profiles for the differently-anchored subsystems with rectangle ratio $R$=2 and strong vertical anchoring
(a) between vertically-aligned substrate regions; (b) between planar-aligned substrate regions.}
\end{center}
\end{figure}

\begin{figure}[h]
\begin{center}
\includegraphics[width=0.5\textwidth]{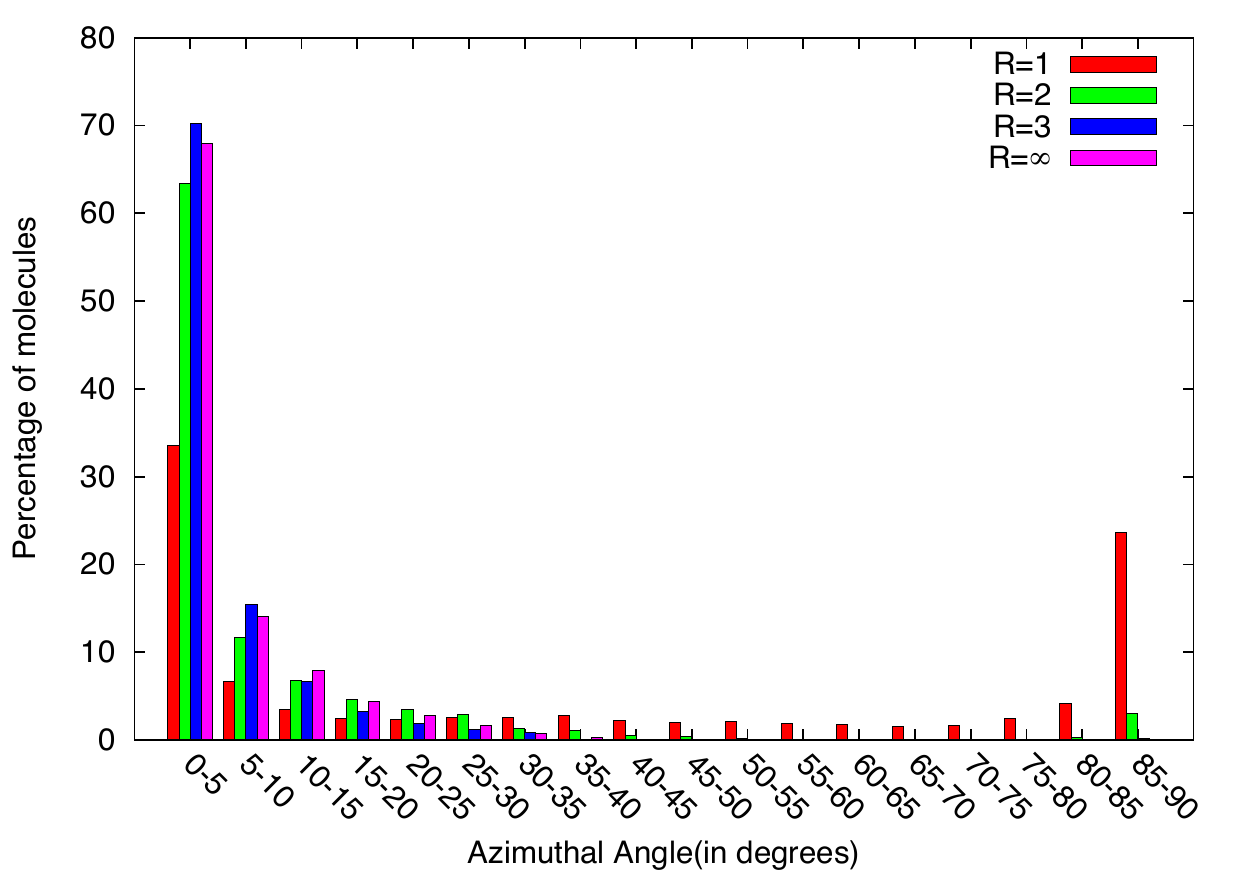}
\caption{\label{histo_comparison_stripe_rectangle}(Online color)Influence of the
pattern anisotropy on the azimuthal angle distribution of near-substrate particles for systems with $k_{s}$=0 and $k_{s}$=3.}
\end{center}
\end{figure}

In order to assess the influence of the rectangle ratio on these small length-scale systems,
an equivalent series of simulations has been performed on systems patterned with $R=L_{x}$/$L_{y}$=2
substrate rectangles.
Corresponding high density snapshots are represented
in Fig.~\ref{snapshot_rectangle_influenceratio_Lz=4K2}.
We note from these that again, in all cases, the induced anchoring lies in the $x$-$z$ plane.
This indicates that fixing the azimuthal angle to coincide with the long edge of the
rectangle patterns is achievable even with $R\simeq$2.
The influence of the ratio $R$ on the azimuthal angle
will be investigated in more detail later in the paper.
\begin{figure}[!h]
\begin{center}
\subfigure[\label{fig:rectcomp1_5_3_homeo}
]{\includegraphics[width=0.5\textwidth]
{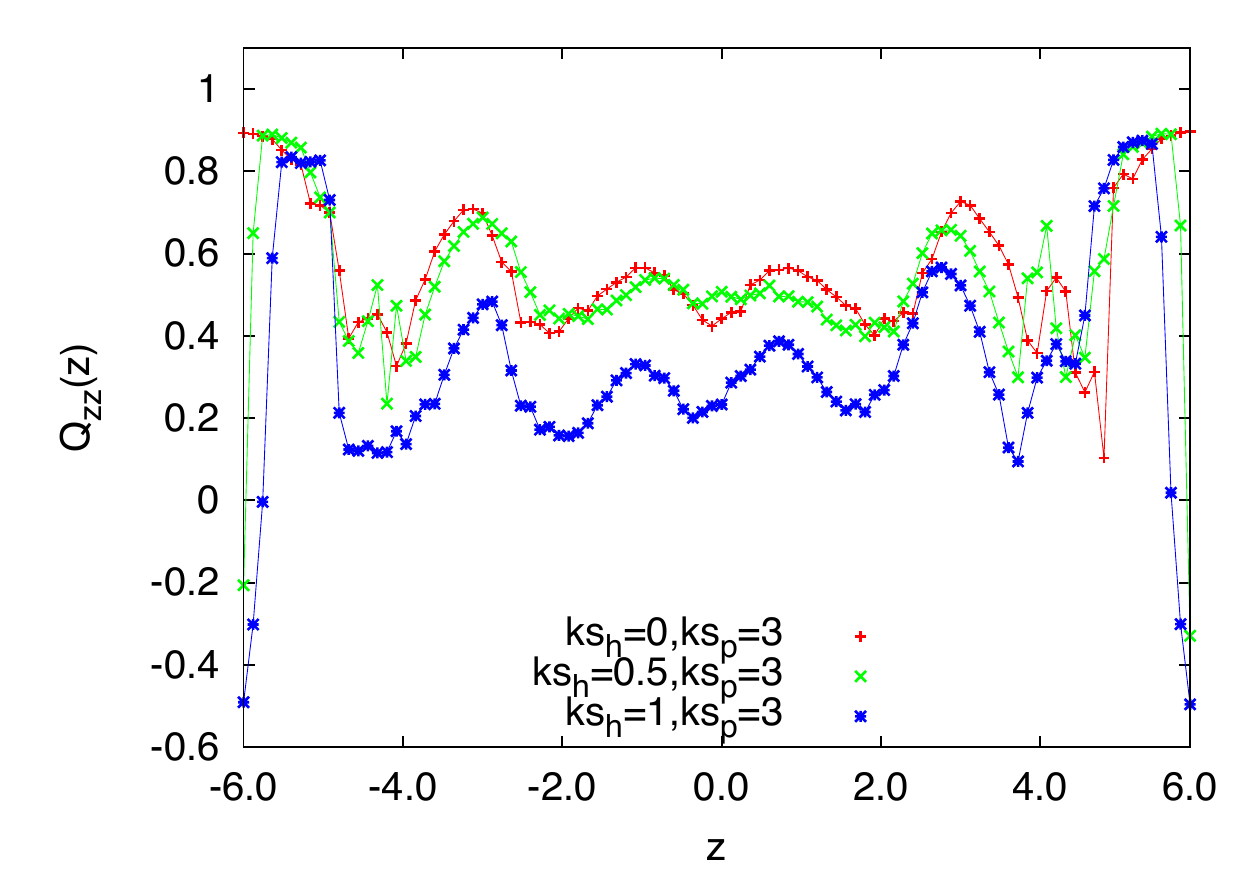}}
\subfigure[\label{fig:rectcomp1_5_3_planar}
]{\includegraphics[width=0.5\textwidth]
{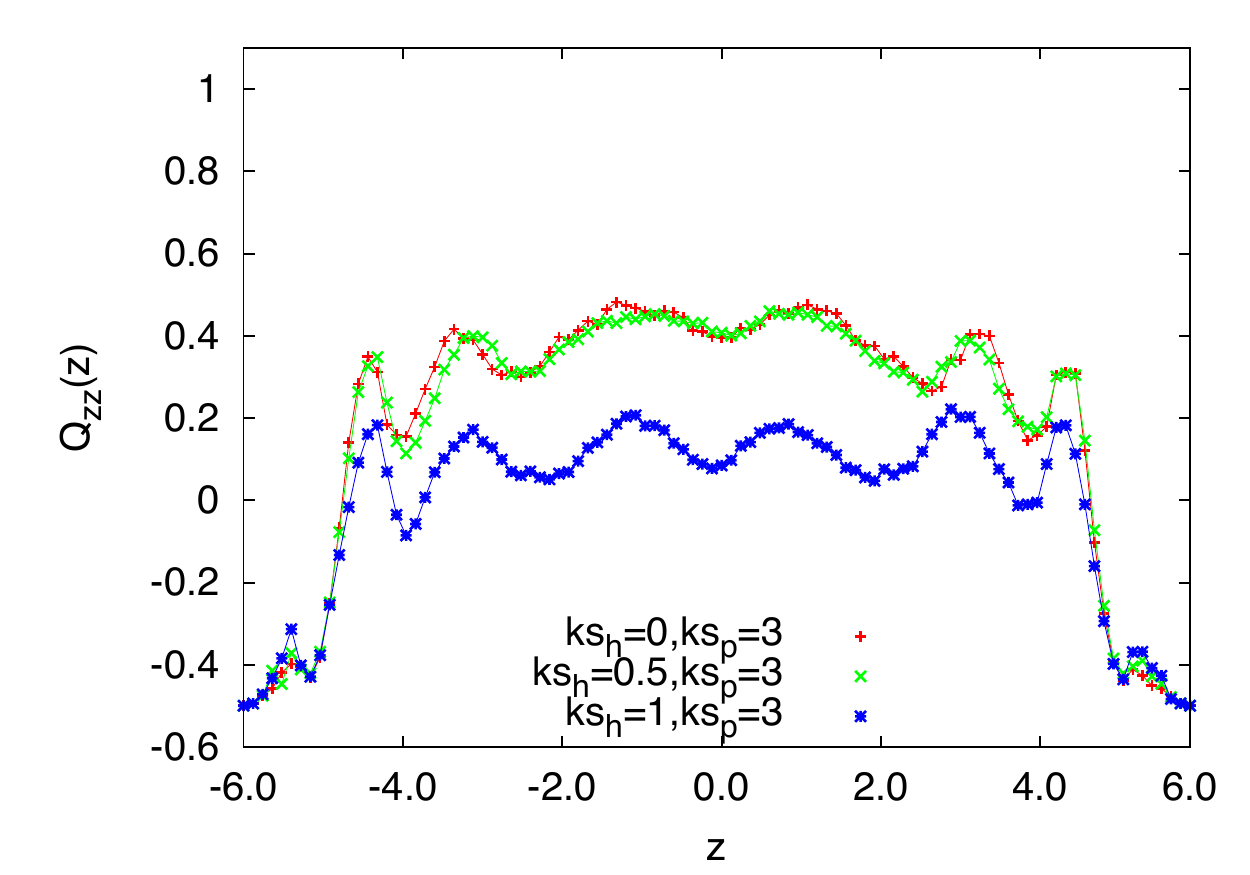}}
\caption{\label{fig:rectcomp1_5_3}(Color online)Influence of $k_{s}$
on the $\rho=0.4$ $Q_{zz}$ profiles for the differently-anchored subsystems with rectangle ratio $R$=2 and strong planar anchoring
(a) between vertically-aligned substrate regions; (b) between planar-aligned substrate regions.}
\end{center}
\end{figure}
\\[0.5cm]
Fig.~\ref{fig:rectcomp1_5_3} shows the high density $Q_{zz}$
profiles of ${R}$=2 systems with strong planar
anchoring regions and varying vertical anchoring strength.
These profiles are more sensitive to the variation of the
vertical surface anchoring parameter than the corresponding $R=3$ were. Specifically, whilst the $R$=2
$Q_{zz}$ profiles are similar for $k_{s}$=0 and
$k_{s}$=0.5, significantly lower $Q_{zz}$ values were found in the central region of the $k_{s}$=1 film.
The corresponding tilt angle shifts are reported in Table~\ref{tiltrect3}.
Fig.~\ref{fig:rectcomp1_9_7} shows the influence of the planar $k_{s}$ parameter on $R$=2 systems with strong vertically aligned regions.
Here, the $Q_{zz}$ values are strongly influenced by the planar $k_{s}$ values, both regions showing increase in $Q_{zz}$ as the planar anchoring component is weakened. This corresponds to the particles lying increasingly normal to the surfaces, as confirmed by the
data given in Tab.~\ref{tiltrect3}.
\newline
\\[0.5cm]
Whilst the qualitative behaviors of the $R=$2 and $R=$3 systems simulated here were very similar, some quantitative differences were determined. From the data reported in Tab.~\ref{tiltrect3}, increase in $L_{x}$/$L_{y}$ was generally associated with a decrease in the tilt angles adopted in both substrate regions. The one exception to this was the substrate combination $k_{s}$=1 $k_{s}$=3 identified above as displaying strong planar character for $R=2$. A further observation is that the sensitivity of tilt angle to substrate conditions was consistently greater for the $R=2$ systems than the $R=3$. It is not appropriate to examine this tilt variation with shape up to the $R=\inf$ limit because such systems exhibit domain bridging rather than a tilted monodomain~\cite{Deck2010}.
\\[0.5cm]
Finally, we assess the effect of the rectangle ratio on the preferred azimuthal angle. In~\cite{Deck2012},
we saw that the azimuthal angle cannot be effectively controlled using square-patterned substrates, degenerate alignment being seen between the two edge directions. Here, though, for both $R$ values investigated, clear orientational pinning was apparent. In order to gauge the strength of this pinning, we plot, in Fig.~\ref{histo_comparison_stripe_rectangle}, azimuthal angle distributions for substrate-region particles on the different pattern types. This confirms a strong departure from the degenerate behaviour seen for square patterns. For $R=2$, while a small but distinct subset of particles aligned along the short rectangle edge, long-edge alignment was dominant. This effect was even more marked for $R=3$ and, indeed, was essentially as strong as that seen for full stripe-patterning.


\section{Experiment}
\begin{figure*}
\begin{center}
{\includegraphics[width=0.4\textwidth] {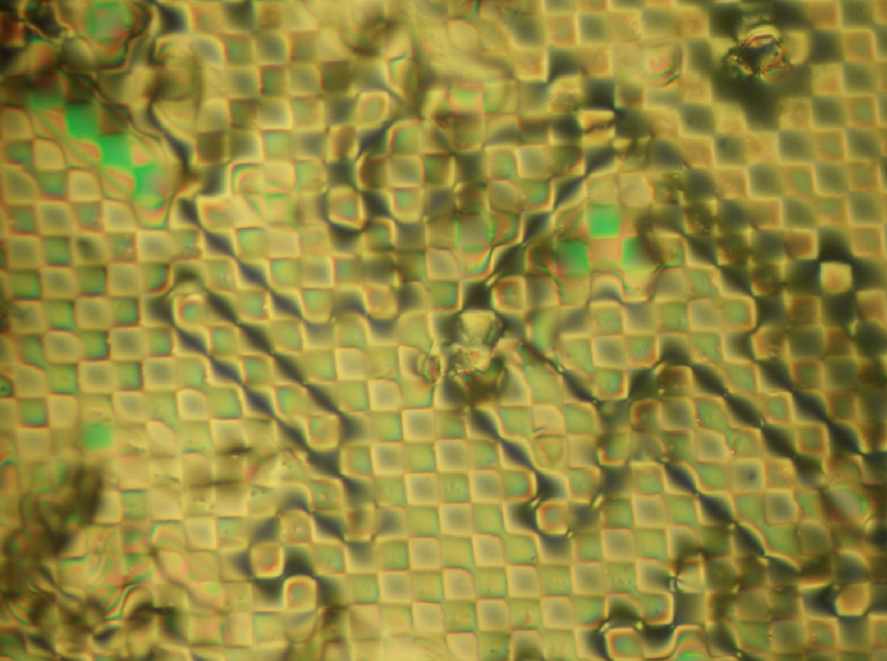}
\includegraphics[width=0.36\textwidth] {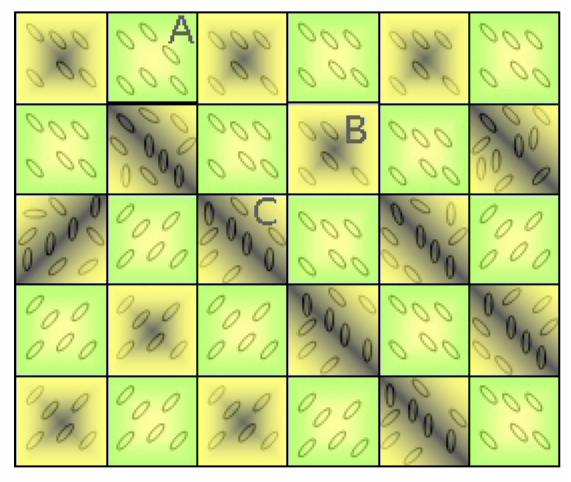}\newline
 }
\caption{\label{fig:square}(Color online) (a)Polarizing microscopy image of 9CB aligned by a COOH/CF3 SAM patterned with a
checkerboard with squares of width 8 $\mu$m. The polarizers
and analyzers are crossed and are parallel to the checkerboard; (b) Schematic showing the bulk director orientation on a square-patterned
substrate. A - PV region, B - VV region with identical neighbors, C - VV region
with different neighboring PV regions. Note the direction of the LC director in a
VV region tries to match that of its neighbors}
\end{center}
\end{figure*}

\begin{figure*}
\begin{center}
\subfigure[\label{b}PM images of 9CB aligned on a COOH/CF$_3$ SAM patterned with a
checkerboard of rectangles, a = 20 $\mu$m, b = 24 $\mu$m. 1) A uniform region of interest 2) 50$^\circ$ anticlockwise sample rotation with crossed polarizers 3) 50$^\circ$ anticlockwise sample rotation with parallel polarizers aligned horizontally.
]{\includegraphics[scale=0.2] {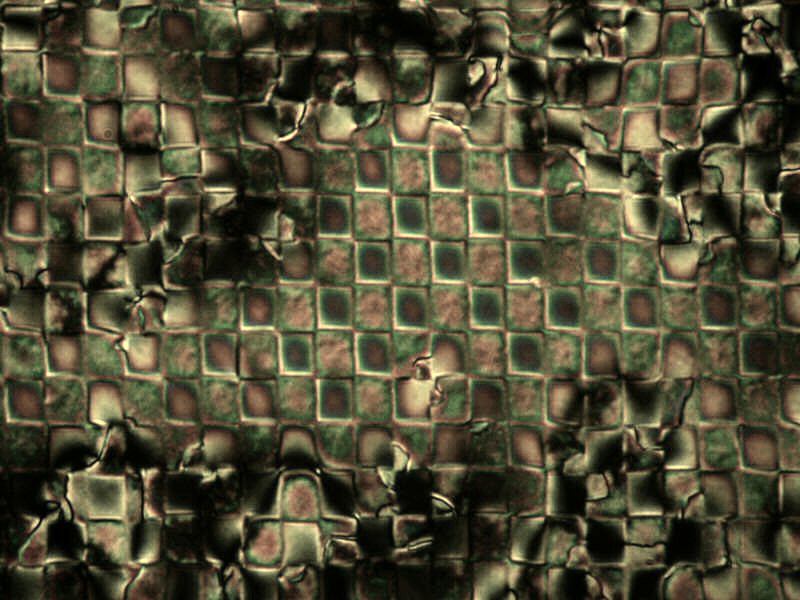}
\includegraphics[scale=0.2] {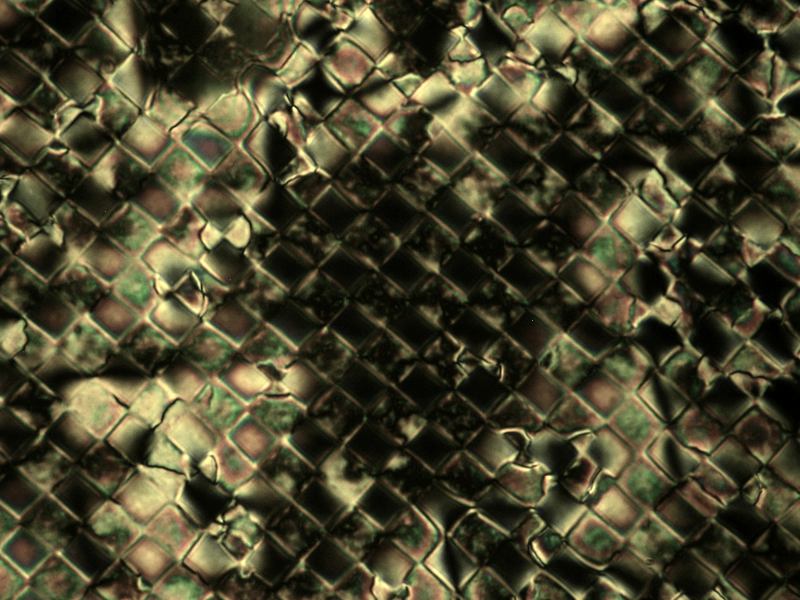}
\includegraphics[scale=0.2] {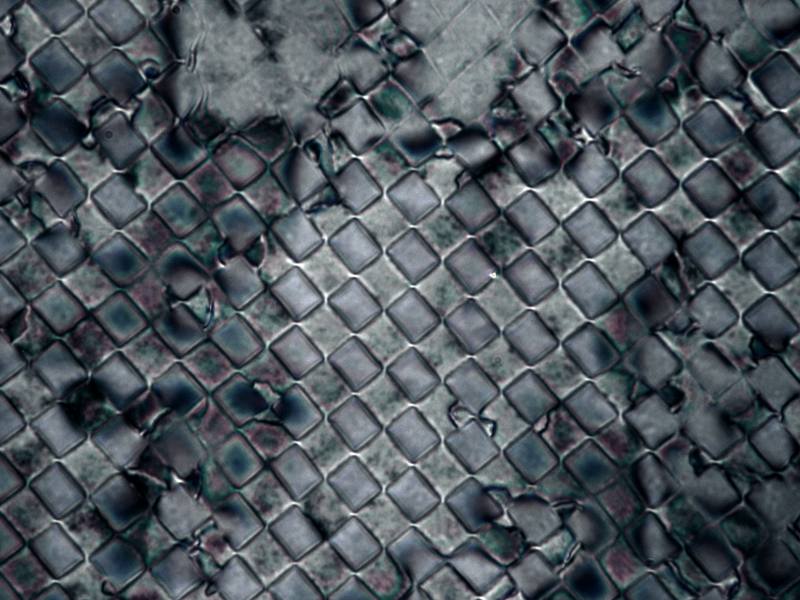}}
\end{center}
\begin{center}
\subfigure[\label{c}PM images of 9CB aligned on a COOH/CF$_3$ SAM patterned with a
checkerboard of rectangles, a = 20 $\mu$m and b = 40 $\mu$m. 1) A uniform region of interest 2) 53$^\circ$ clockwise sample rotation with crossed polarizers 3) 53$^\circ$ clockwise rotation with parallel polarizers aligned horizontally.
]{\includegraphics[scale=0.2] {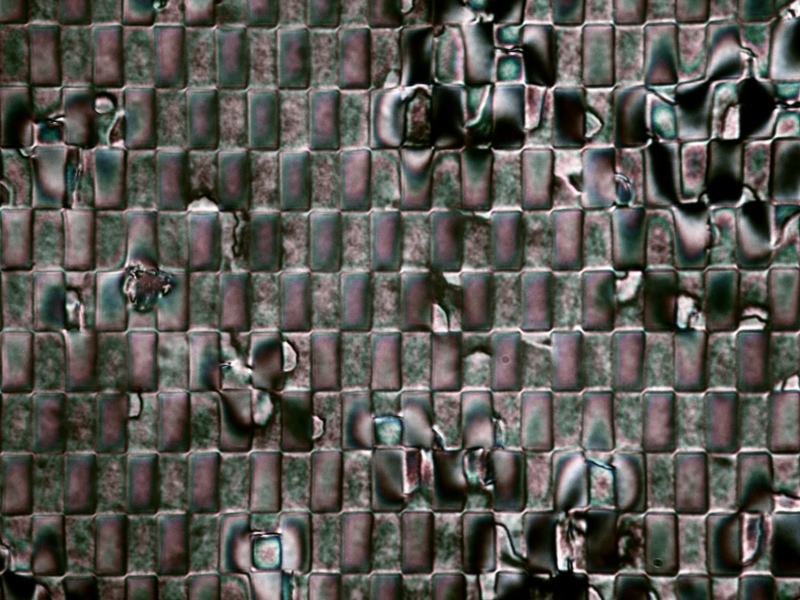}
\includegraphics[scale=0.2] {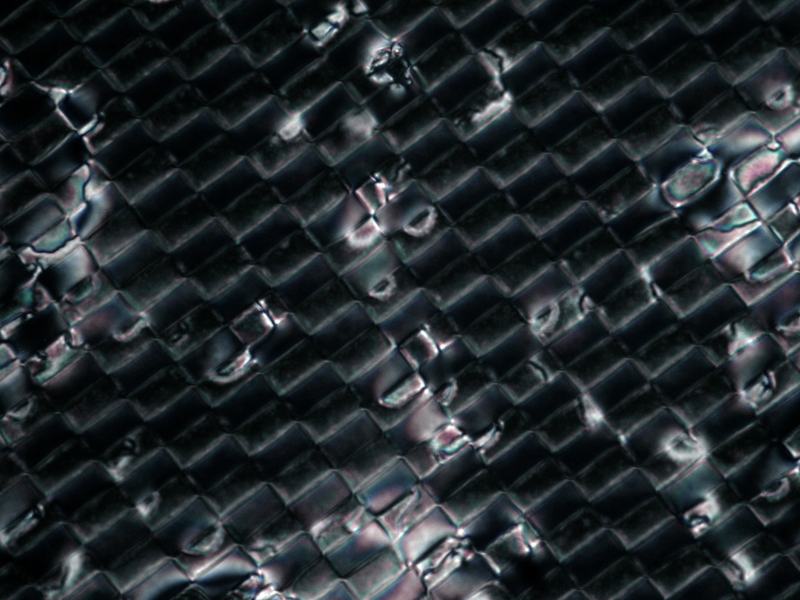}
\includegraphics[scale=0.2] {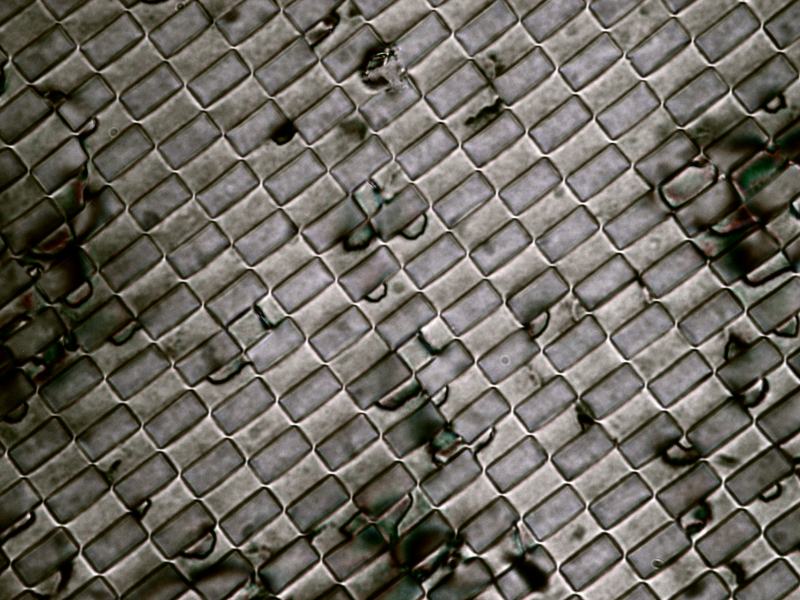}}
\caption{\label{fig:Polarizing-microscopy-images}(Color online)Polarizing microscopy images
of 9CB on a rectangular patterned surface.}
\end{center}
\end{figure*}
To determine the effect of rectangular substrate patterning at larger length-scales,
a set of experimental cells was prepared using the microcontact printing
technique described in \cite{Bramble2007}. In these, SAMs
of -COOH and -$\text{CF}_{3}$ terminated alkanethiols were used to
promote planar and vertical alignment, respectively, in the different pattern regions. Unlike the simulated systems, however, only one patterned substrate was used in these cells; the second substrate was always
prepared with a uniform vertical-aligning treatment. We do not expect the uniform vertical surface to significantly change the azimuthal alignment since the MC simulations show that the elastic distortion is strongly confined to the surface in these systems. Moreover, the uniform vertical treatment for symmetry reasons cannot itself favor any particular azimuthal orientation. Consequently, these cells comprised both
vertical-vertical (VV) and planar-vertical (PV) sub regions. The cells were
prepared with a nominal thickness $23\mu m$, filled with the nematic
material 9CB in the isotropic phase and cooled into the nematic phase
while being observed with a polarizing microscope.

Fig.~\ref{fig:square} shows a polarizing microscopy
image of a typical cell, one surface of which was patterned with squares of 8$\mu m$ width (i.e. a 16
$\mu m$ period). For this system, the maximum transmission in the planar regions occurred when the checkerboard was orientated parallel to the polarizers. From this, we deduce that the director had a component aligned
diagonally across the planar-aligned squares. There are two different degenerate states for this,
corresponding to the two opposite diagonals, which are optically
identical. Nonetheless, optical consequences of splay and twist in the VV regions reveals where boundaries between the differently oriented PV domains
lie: at an interface between these two the azimuthal angle $\phi$ rotates gradually by $\pi/2$ from one side to the other, so along the diagonal of the square, $\phi$ averages to be parallel to one of the polariser directions and hence appears dark. The VV regions are splayed and twisted in the bulk of the cell due to their requirement to accommodate the splayed state in the neighboring PV regions. When
a VV region is surrounded by four PV regions of the same hybrid aligned nematic
(HAN) configuration, its director points in the same diagonal plane as do those of its neighbors.
However, when it is surrounded by PV regions with differing HAN configurations,
the VV region is not able to match its boundary requirements with a simple HAN profile.
Instead, a twist component develops, which can been seen optically
as a darker VV square. As these dark VV squares only arise where there are neighbors with
different PV region alignments, they form a border around domains of the
degenerate HAN states. This figure shows as well a schematic illustrating these different states
and their relationship to the microscopy image.
This diagonal alignment behavior in VV regions is predicted by continuum model calculations~\cite{Deck2012} if the polar anchoring energy promoted by the surface is rather weak.

A second set of cells was prepared with rectangular patterns of different
aspect ratios; otherwise, parameters and conditions were equivalent to those
used for the square-patterned cell. Microscopy images are shown in Fig.~\ref{fig:Polarizing-microscopy-images}(a) and (b) and the measured azimuthal alignment angles are displayed in Table \ref{tab:Observed-alignment-of}; these were obtained as for the square pattern by rotating the sample under the microscope so as to maximize the extinction between crossed polarisers.

When the symmetry of square checkerboards was broken in this way, a number of alignment changes were observed. We still have two degenerate azimuthal states seen for the regular squares, but the azimuthal orientation in the centre is no longer at $45^\circ$.  On rotation between crossed polarisers, dark states were observed for anti-clockwise (negative) and clockwise (positive) rotations. Defining a positive angle $\phi$ from the long axis of the rectangle, dark states occur at rotations of -$\phi^\circ$, 90-$\phi^\circ$ for one state ($\phi$) and -(90-$\phi$)$^\circ$, 90+$\phi^\circ$ for the other state (-$\phi$). However optical degeneracy also has to be considered, as the same configurations could be equally explained by azimuthal angles of 90-$\phi^\circ$ and -(90-$\phi$)$^\circ$. Here we use the fact that, when observing with parallel polarisers set horizontally, there is no contrast between the PV regions and VV regions if the horizontal components of refractive index in the HAN and vertical states are equal. This allows us to correctly identify the azimuthal angles. Optical compensators or direct imaging with fluorescent confocal polarising microscopy (FCPM) could also be used to distinguish between these optically degenerate states.

For an aspect ratio of 1.2 at $40^\circ$ measured from the vertical. Dark states can be seen on clockwise rotation to $40^\circ$ and anticlockwise rotation to $50^\circ$, the latter shown in fig. \ref{fig:Polarizing-microscopy-images}(a)(2). Parallel polarizer images also show the difference between the bistable states \ref{fig:Polarizing-microscopy-images}(a)(3). The extinction angle is intriguingly close to $\arctan(1/1.2) = 39.8^\circ$, the angle of the director if pointing from the center of the rectangle to a corner. Additional states are also observed, as shown in Figure \ref{fig:Polarizing-microscopy-images}(a)(3), there are regions where the director is parallel to the long axis of the rectangle. These regions have a bright state on rotation to $45^\circ$ between crossed polarizers.
Patterns with an increasing ratio up to 2 were printed, the final example of which can be seen in figure \ref{fig:Polarizing-microscopy-images}(b) with similar behavior to the aspect ratio $1.2$ case.
The observed alignment angles are summarized in Table \ref{tab:Observed-alignment-of}. Although the MC simulation predicts a different alignment direction to that observed experimentally, there is a qualitative agreement between the two in that the tendency of the nematic is to align with the longer side of the rectangle with increasing aspect ratio.
\begin{table}
\begin{centering}
\begin{tabular*}{3.4in}{@{\extracolsep{\fill}}ccc>{\centering}p{1in}}
\hline
Width ($\mu$m) & length ($\mu$m) & Aspect Ratio & Azimuthal angle from long axis ($^\circ$)\tabularnewline
\hline
\hline
20 & 20 & 1 & 45\tabularnewline
20 & 24 & 1.2 & 40\tabularnewline
20 & 28 & 1.4 & 36\tabularnewline
20 & 32 & 1.6 & 32\tabularnewline
20 & 36 & 1.8 & 29\tabularnewline
20 & 40 & 2 & 27\tabularnewline
\hline
\end{tabular*}
\par\end{centering}

\caption{\label{tab:Observed-alignment-of}Observed alignment of 9CB on rectangle
patterned surfaces. }
\end{table}

\pagebreak

\section{Continuum Model}

To reconcile the apparently contradictory observations from our MC simulations and experiments, We now examine the rectangle pattern using a continuum theory approach following
the procedure described in our previous paper\cite{Deck2012}. Our objective
is to determine the director field
\begin{equation}
\mathbf{\hat{n}}(\vec{x})=\left(\cos\theta\sin\phi,\cos\theta\cos\phi,\sin\theta\right)\label{eq:director}
\end{equation}
that minimizes the free energy consisting of the Frank energy and
the surface energy
\begin{widetext}
\begin{eqnarray}
F=\frac{1}{2}\int d^{3}x\ K_{1}\left(\nabla\cdot\mathbf{\hat{n}}\right)^{2}+K_{2}\left[\mathbf{\hat{n}}\cdot(\nabla\times\mathbf{\hat{n}})\right]^{2}+K_{3}\left|\mathbf{\hat{n}}\times(\nabla\times\mathbf{\hat{n}})\right|^{2}+\int_{s}\text{d}S\ g(\vec{\mathbf{n}},\vec{\mathbf{n}}_{0})
\label{eq:Frank}
\end{eqnarray}
\end{widetext}
within two simplifying assumptions: that $K_{1}=K_{3}\neq K_{2}$ with $\tau=K_2/K_1$;
and also that only variations in $\theta$ are considered, i.e. that
the director is confined everywhere to a single plane. The first assumption
is approximately justified for common nematic mesogens, the second
is reasonable since, as evidenced by Figs.~\ref{fig:Polarizing-microscopy-images}
a) and b), the patterned surface promotes variation in the
polar coordinate only and places no restriction on the azimuthal
orientation. Here, the coordinate system is scaled by the thickness
of the liquid crystal layer $L_z$, so that the periods of the pattern
in the $x$ and $y$ directions, $L_{x}$ and $L_{y}$,
are dimensionless quantities.

Here, since $L_{x}\neq L_{y}$, we require a solution of
the form
\begin{widetext}
\begin{eqnarray}
\theta(x,y,z)=\theta_{0}+\sum_{n=-\infty}^{\infty}\sum_{m=-\infty}^{\infty}\frac{1}{\sqrt{L_{x}L_{y}}}\left(A_{nm}e^{-\nu_{nm}z}+B_{nm}e^{\nu_{nm}z}\right)\exp\left[i2\pi\left(nx/L_{x}+my/L_{y}\right)\right].\label{eq:solution}
\end{eqnarray}
\end{widetext}
The Euler-Lagrange equation\cite{Deck2012}
\begin{widetext}
\begin{eqnarray}
\left(\tau\cos^{2}\phi+\sin^{2}\phi\right)\frac{\partial^{2}\theta}{\partial x^{2}}+\left(\tau\sin^{2}\phi+\cos^{2}\phi\right)\frac{\partial^{2}\theta}{\partial y^{2}}+(1-\tau)\sin(2\phi)\frac{\partial^{2}\theta}{\partial x\partial y}+\frac{\partial^{2}\theta}{\partial z^{2}}=0\label{eq:EulerLangrange}
\end{eqnarray}
\end{widetext}
is satisfied if the $\nu_{nm}$ are chosen
\pagebreak
\begin{widetext}
\begin{eqnarray}
\nu_{nm}=\frac{\pi}{L_{x}L_{y}}\sqrt{2(\tau+1)\left(L_{x}^{2}m^{2}+L_{y}^{2}n^{2}\right)-2(\tau-1)\left[2L_{x}L_{y}mn\sin(2\phi)+\cos(2\phi)\left(L_{x}^{2}m^{2}-L_{y}^{2}n^{2}\right)\right]}.
\end{eqnarray}
\end{widetext}
The remaining coefficients are obtained from the weak anchoring boundary
conditions, which for the harmonic anchoring potential
\begin{equation}
g_{H}(\theta,\theta_{e})=\frac{W_{\theta}}{2}(\theta-\theta_{e})^{2}\label{eq:harmonic}
\end{equation}
are
\begin{figure}
\begin{center}
\includegraphics[width=0.5\textwidth]{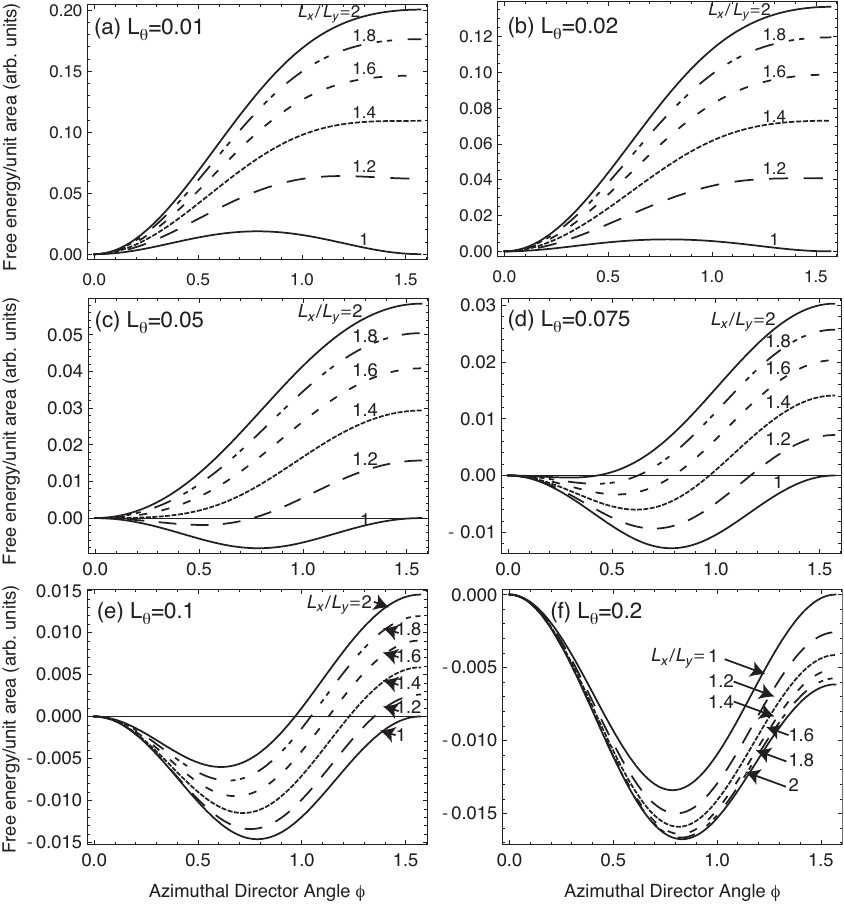}
\caption{\label{fig:FreeEnergyPhi}Free energy per unit area as a function
of $\phi$, plotted for various values of $L_{y}/L_{x}$. }
\end{center}
\end{figure}
\begin{equation}
\pm L_{\theta}\frac{\partial\theta}{\partial z}+\theta=\theta_{e}\label{eq:Robin}
\end{equation}
where $\theta_{e}(x,y)$ is the spatially varying easy axis promoted by the pattern and the sign corresponds to the direction of the surface normal
pointing out of the liquid crystal layer at the appropriate boundary
and where the dimensionless parameter associated with polar anchoring
$L_{\theta}$ is
\begin{equation}
L_{\theta}=\frac{K_{1}}{W_{\theta}L_z}.\label{eq:AnchoringLength}
\end{equation}
As a guide to interpreting this number, for a cell thickness $L_z=10\mu\text{m}$
and typical unpatterned values of $W_{\theta}=10^{-5}\text{J}\text{m}^{-2}$ and
$K_{1}\approx10pN$ (5CB) the value of $L_{\theta}=0.01$.

By inserting (\ref{eq:solution}) into (\ref{eq:Robin}) and performing
routine calculations, the coefficients $\theta_{0}$, $A_{nm}$ and
$B_{nm}$ are obtained

\begin{eqnarray}
\theta_{0} & = & \pi/4\nonumber \\
A_{nm} & = & \frac{e^{\nu_{nm}}c_{nm}}{L_{\theta}\nu_{nm}(e^{\nu_{nm}}-1)+(e^{\nu_{nm}}+1)},\\
B_{nm} & = & \frac{c_{nm}}{L_{\theta}\nu_{nm}(e^{\nu_{nm}}-1)+(e^{\nu_{nm}}+1)},\label{eq:ABcoeff}
\end{eqnarray}
where the $c_{nm}$ are the Fourier coefficients of the easy axis
profile $\theta_{0}(x,y)$ at the $z=0$ and $z=1$ surfaces respectively,
\begin{equation}
c_{nm}=d_{nm}=\begin{cases}
-\frac{\sqrt{L_{x}L_{y}}}{\pi nm} & n,m\ \text{odd}\\
0 & \text{otherwise}
\end{cases}.\label{eq:cdcoeff}
\end{equation}
Having determined the solution as above, the free energy may be evaluated
by inserting \ref{eq:solution} into (\ref{eq:Frank}) and performing necessary integrations.
The bulk energy is
\begin{widetext}
\begin{eqnarray}
F_{b} & = & \sum_{nm}\frac{\pi^{2}}{L_{x}^{2}L_{y}^{2}\nu_{nm}}\left[\left(A_{nm}^{2}e^{-\nu_{nm}}+B_{nm}^{2}e^{+\nu_{nm}}\right)\sinh(\nu_{nm})+2A_{nm}B_{nm}\nu_{nm}\right]\times\nonumber \\
&  & \ \ \ \ \ \times\left\{ (1+\tau)\left(L_{x}^{2}m^{2}+L_{y}^{2}n^{2}\right)+(1-\tau)\left[\cos(2\phi)\left(L_{x}^{2}m^{2}-L_{y}^{2}n^{2}\right)+2L_{x}L_{y}mn\sin(2\phi)\right]\right\} +\nonumber \\
&  & +\sum_{nm}\frac{1}{2}\nu_{nm}\left[\left(A_{nm}^{2}e^{-\nu_{nm}}+B_{nm}^{2}e^{+\nu_{nm}}\right)\sinh(\nu_{nm})-2A_{nm}B_{nm}\nu_{nm}\right].\label{eq:bulkenergy}
\end{eqnarray}
\end{widetext}
The surface energy (for each surface) is
\begin{equation}
F_{s}=\pi^{2}L_{x}L_{y}/16+\frac{1}{L_{\theta}}\sum_{nm}(A_{nm}+B_{nm})(A_{nm}+B_{nm}-2c_{nm}).\label{eq:surf}
\end{equation}

The free energy per unit area as a function of $\phi$ is plotted
in Fig.~\ref{fig:FreeEnergyPhi}
for different values of $L_{\theta}$ and $L_{y}/L_{x}$ with
$\tau=1/2$ and $L_{x}=1$. For the strongest anchoring depicted, where $L_{\theta}=0.05$, it is apparent that a very slight difference in $L_{x}$ and $L_{y}$ is sufficient to break the degeneracy of the configurations
aligned along the $x$ and $y$ axes respectively; once $L_{y}/L_{x}\gtrsim1.5$
there is no stable state aligned along the $x$ axis.

Rectangle-patterned surfaces are surfaces of adjustable azimuthal
anchoring energy: they may be thought of as promoting an effective
azimuthal anchoring potential~\cite{Harnau:2007p2047}, the strength of which is varied
by modest changes in the aspect ratio of the rectangles. From the
plot in Fig.~\ref{fig:FreeEnergyPhi}(a), it is apparent that over the
range of aspect ratios $1\text{\textemdash}2$, the effective azimuthal
anchoring energy varies by a factor of roughly $5$. The controllability
is, however, contingent on the ability of the pattern to deform the
nematic as measured by the anchoring strength $L_{\theta}$. Free energy profiles as a function of $\phi$ for increasing values of $L_{\theta}$, corresponding to weaker anchoring, are displayed in Fig.~\ref{fig:FreeEnergyPhi}(b-f). As $L_{\theta}$ increases, the anchoring transition described in our previous paper~\cite{Deck2012} occurs (Fig.~\ref{fig:FreeEnergyPhi}c)) whereby the preferred azimuthal orientation, indicated by the position of the minimum, is no longer along the length
or breadth of the square pattern, but lies roughly along the diagonal. For the largest value of $L_{\theta}=0.2$ plotted, the depth of the effective azimuthal anchoring potential is significantly smaller than for the $L_{\theta}=0.05$ case.

The variation of the preferred azimuthal orientation as a function of aspect ratio also depends strongly on the anchoring parameter around the anchoring transition as may be seen in Fig.~\ref{fig:FreeEnergyPhi}(c-e), while both the edge- (Fig.~\ref{fig:FreeEnergyPhi}(a,b)) and diagonally-aligned states (Fig.~\ref{fig:FreeEnergyPhi}(f)) are insensitive to the aspect ratio of the pattern . The experiment described in the previous section took place in this diagonal regime and so the observed alignment angle as a function of aspect ratio can be used to estimate the anchoring parameter; from Tab.~\ref{tab:Observed-alignment-of} we obtain a value of $L_{\theta}\sim$0.08. Here, the apparent contradiction between the Monte Carlo and experimental results is resolved by the continuum prediction that there exist two different anchoring regimes characterized by the anchoring parameter $L_{\theta}$; the MC simulations and experiment have illuminated the two regimes. In all models the tendency is for the director to align with the long axis of the rectangles.

\section{Conclusions}
In this work, a surface prepared with a chessboard-like pattern of alternating vertical and planar rectangles has been shown to azimuthally align an adjacent nematic liquid crystal. Moreover, control of the alignment easy axis has been demonstrated in both azimuthal and zenithal coordinates achieved by adjusting the design parameters of the pattern, i.e. the lengthscale, aspect ratio and relative anchoring strength of the rectangles.
\\[0.5cm]
Monte Carlo simulations of hard particles predict that the preferred alignment direction is along the long edges of the rectangles and that only modest aspect ratios $\sim2-3$  are required to break the bistable alignment previously observed in square-patterned systems\cite{Deck2012}. Hence, rectangle patterns resemble the striped system extensively studied \cite{Atherton:2006p31,Deck2010,Atherton:2010p3029} except that the bulk alignment is a monodomain and no bridging behavior, where the nematic follows the pattern throughout the film, was observed; there is only weak modulation in the polar angle at the film center, in agreement with that predicted by continuum theory. Control of the tilt angle over a range of $\sim20^{\circ}-40^{\circ}$  was observed by running a sequence of MC simulations adjusting the relative anchoring strength of the vertical and planar regions. Simulations with increasing $L_{x}/L_{y}$ had a distribution of particles increasingly strongly peaked around the alignment direction, suggesting that the azimuthal anchoring strength increases as a function of aspect ratio.
\\[0.5cm]
Experimental observations of a nematic aligned on rectangle surfaces prepared by chemical patterning reveal a different behavior to that seen in the Monte Carlo simulation: while for increasing aspect ratio the azimuthal alignment direction indeed becomes more oriented toward the long edge of the rectangles, the alignment direction is along the diagonals and not the edges.
\\[0.5cm]
The disparity is resolved by our continuum model which predicts an anchoring transition from diagonal to edge alignment if the anchoring parameter $L_{\theta}\lesssim0.05$ . The thin film, only a few particles thick, used in the MC simulation is inside the edge-aligned strong anchoring regime. By fitting the observed variation of the azimuthal anchoring direction as a function of aspect ratio, we infer that the experimental anchoring parameter was $L_{\theta}\sim0.08$ . The proximity of this value to the critical value $L_{\theta}\sim0.05$  suggests that with a suitable choice of materials, cell thicknesses and patterns that the transition out to be observable in a future experiment.
\\[0.5cm]
The key parameter in these systems $L_{\theta}\sim0.05$, introduced in eqn.(\ref{eq:AnchoringLength}), is simply the ratio of the bulk splay elasticity to the product of the unpatterned polar anchoring strength and the film thickness. Thus, it is apparent that the edge-aligned behaviour observed in the MC simulation part of our study was {\em not} dictated by their small length-scale - in fact, the small $d$ of the simulated systems acted to promote the diagonal alignment regime. That the simulations remained firmly in the small $L_{\theta}\sim0.05$ regime is, then, a consequence of the very high anchoring coefficients pertaining to all models described in terms of monodisperse anisotropic particles adsorbed at planar substrates.
\\[0.5cm]
Previous studies\cite{Spencer_care} have shown that the fine-tuning of surface anchoring, typically achieved by selection of materials and preparation conditions, is necessary to optimize the performance of advanced electro-optic devices. Patterned substrates of the type considered here, however, have the potential to offer a more convenient and well-controlled route to achieving such optimisation, since all of the surface anchoring parameters, i.e. the easy axis, polar and azimuthal anchoring energies, can be controlled by adjusting the design features of the pattern. In such systems, the emergent polar and azimuthal anchoring parameters can be viewed as resulting from a convolution of the basic material parameters and the imposed patterning. The conventional continuum description, thus, corresponds to an effective integration of a potential spectrum of various surface-printed features on length-scales ranging from nanometres to microns.

\begin{acknowledgments}
This work was supported by the Engineering and Physical Research
Council, Grant No GR/S59833/01. We acknowledge useful conversations with Chris Care, Tim Spencer and Paulo Teixeira which have been beneficial to our understanding of the systems studied here. CA-D and DJC performed the MC simulations, JPB performed the experiments with Jim Henderson and Stephen D Evans, TJA constructed the continuum theory; CA-D, DJC and TJA prepared the manuscript with advice from JPB.
\end{acknowledgments}

\end{document}